\documentstyle[prl,aps,preprint,amssymb,epsfig]{revtex}

\begin{document}

\def\d{{\rm d}}
\def\e{{\rm e}}
\def\O{{\rm O}}
\def\half{\mbox{$\frac12$}}
\def\eref#1{(\protect\ref{#1})}
\def\etal{{\it{}et~al.}}

\title{Efficient Monte Carlo algorithm and high-precision results for
percolation}
\author{M. E. J. Newman$^1$ and R. M. Ziff$^2$}
\address{$^1$Santa Fe Institute, 1399 Hyde Park Road, Santa Fe, NM 87501}
\address{$^2$Dept.\ of Chemical Engineering, University of Michigan,
Ann Arbor, MI 48109--2136}
\maketitle

\begin{abstract}
  We present a new Monte Carlo algorithm for studying site or bond
  percolation on any lattice.  The algorithm allows us to calculate
  quantities such as the cluster size distribution or spanning probability
  over the entire range of site or bond occupation probabilities from zero
  to one in a single run which takes an amount of time scaling linearly
  with the number of sites on the lattice.  We use our algorithm to
  determine that the percolation transition occurs at $p_c =
  0.59274621(13)$ for site percolation on the square lattice and to provide
  clear numerical confirmation of the conjectured $4/3$-power
  stretched-exponential tails in the spanning probability functions.
\end{abstract}

\vfill
\begin{center}
Published as {\it Phys.\ Rev.\ Lett.} {\bf85}, 4104--4107 (2000).
\end{center}

\newpage

Percolation~\cite{SA91} is one of the best studied problems in statistical
physics, both because of its fundamental nature and because of its
applicability to a wide variety of different systems.  Percolation models
have been used as a representation of resistor networks~\cite{ARC85},
forest fires~\cite{Henley93}, epidemics~\cite{MN00}, biological
evolution~\cite{JBHS94}, and social influence~\cite{SWAJS00}, as well as, of
course, percolation itself.  The word percolation appears in the title of
almost four thousand physics papers in the last quarter of a century.

Numerical studies of percolation are straightforward by comparison with
many simulations in statistical physics because no Markov process is needed
to perform importance sampling.  One can generate a single correct sample
from the ensemble of possible states of a site (bond) percolation model on
any lattice simply by populating each of the sites (bonds) of that lattice
independently with occupation probability $p$.  Typically one then finds
all the connected clusters of occupied sites (bonds) in the resulting
configuration using either depth-first or breadth-first search, and uses
this information to calculate some observable of interest, such as average
or largest cluster size.  An extension of this method is the well-known
Hoshen--Kopelman algorithm~\cite{HK76}, which allows one to find all the
clusters while storing the state of only a small portion of the lattice at
any time.  Other numerical algorithms have been developed to answer
specific questions about percolation models, such as the hull-generation
algorithm~\cite{ZCS84,Grassberger92}, which can tell us whether a cluster
exists which spans a square lattice with open boundary conditions without
actually populating all the sites of that lattice first.

All of these algorithms have one feature in common: they tell us about the
properties of the system for one specified value of $p$ only.  In most
cases one would like to know about the properties of the system over a
range of values anywhere up to the entire domain $0\le p\le1$.  Although
$p$ can in theory take any real value in this range, we need not, on a
system of finite size, study an infinite number of values of $p$ to answer
a question about that system with arbitrary precision.  In fact, on a
system of $N$ sites, we need measure an observable $Q$ only for systems
having a {\em fixed\/} number $n$ of occupied sites (or bonds).  We will
refer to this as the ``microcanonical ensemble'' for the percolation
problem.  If we can measure the values $Q_n$ of our observable for all
$0\le n\le N$ (or the equivalent range for bond percolation), then we can
find the value $Q(p)$ for the more common ``canonical ensemble'' for any
value of $p$ by convolution with a binomial distribution:
\begin{equation}
Q(p) = \sum_n \left({N\atop n}\right) p^n (1-p)^{N-n} Q_n.
\label{convolve}
\end{equation}
Both depth-first and breadth-first searches take time $\O(N)$ to construct
all clusters, and since there are $\O(N)$ possible values of the number of
occupied sites or bonds, it is therefore possible to calculate $Q(p)$ over
the entire range of $p$ in time $\O(N^2)$.  The hull-generating algorithm
can perform the same calculation marginally faster, in time $\O(N^{15/8})$,
but is, as mentioned above, restricted to measuring only certain
observables such as the existence (or not) of a system-spanning cluster.
Histogram interpolation methods~\cite{Hu92} can reduce the time taken for a
general measurement to $\O(N^{3/2})$, at the cost of a reduction in
numerical precision, while the position of the percolation point can be
found in time $\O(N\log N)$, by performing a binary search among the $N$
possible values of $n$~\cite{SA91,HA96}.

In this paper we present a new algorithm which can find the value of a
quantity or quantities over the entire range of $p$ from zero to one in
time $\O(N)$---an enormous improvement over the simple $\O(N^2)$ algorithm
described above.  As a corollary, the algorithm can also find the position
of the percolation point in time $\O(N)$, since one can consider the
existence (or not) of a spanning cluster to be the observable of interest.
Our algorithm calculates the value of the quantity or quantities of
interest for all values of $n$ in the microcanonical ensemble described
above and the value in the canonical ensemble can then be calculated by
employing Eq.~\eref{convolve}.  We describe the algorithm first for the
bond percolation case, which is slightly simpler than site percolation.

\begin{figure}
\begin{center}
\psfig{figure=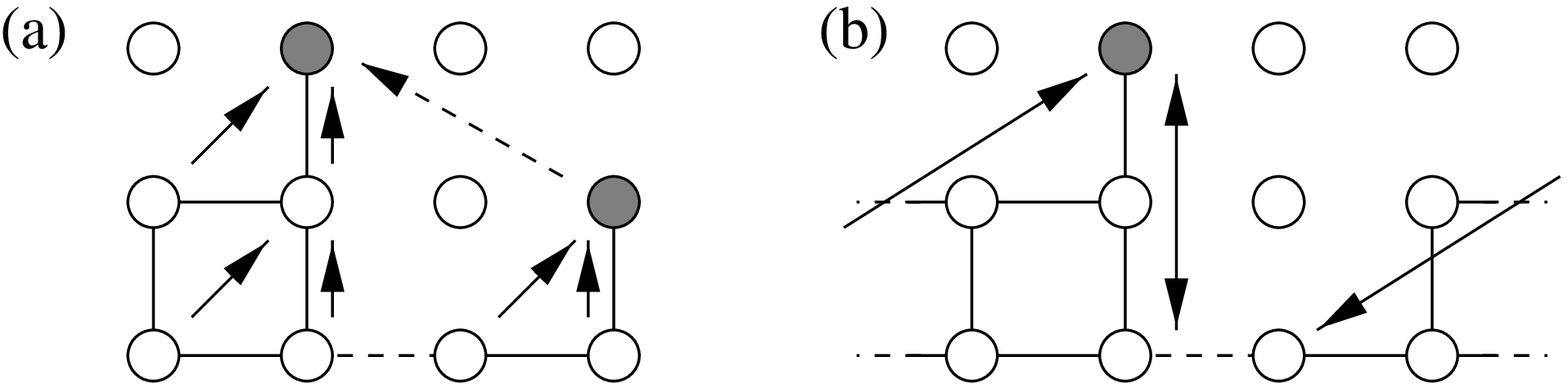,width=5in}
\end{center}
\caption{Two adjacent clusters in a bond percolation system.  (a)~The
arrows represent pointers and the shaded sites are root nodes.  (b)~The
difference in displacements (double-headed arrows) between the two sites
and the root node of a cluster can be used as a criterion for detecting the
onset of percolation.}
\label{tree}
\end{figure}

The basic idea behind our algorithm is the following.  We start with a
lattice in which no bonds are occupied, and hence every site is a separate
cluster.  Each of these single-site clusters is given a unique label
(e.g.,~a positive integer) by which we identify it.  We then fill in bonds
on the lattice in random order.  When a bond is added to the lattice it
either connects together two sites which are already members of the same
cluster---in which case we need do nothing---or it connects two sites which
are members of two different clusters.  In this second case we must change
the labels of one of the clusters to reflect the fact that the new bond has
amalgamated the two.  In order to accomplish this efficiently, we store the
clusters using a tree-structure in which one site in each cluster is chosen
to be the ``root node'' of that cluster and contains the cluster label.
All other sites in the cluster possess pointers which point either to the
root node, or to another site in the cluster, such that by following a
succession of such pointers one can get from any site to the root node.
This scheme is illustrated for the case of the square lattice in
Fig.~\ref{tree}a.  (A similar scheme is used in the Hoshen--Kopelman
algorithm~\cite{HK76}.)  Clusters can now be efficiently amalgamated simply
by adding a pointer from the root node of one to the root node of the other
(dotted arrow in the figure), thereby making the former a sub-tree of the
latter~\cite{note1}.

Our algorithm consists of repeatedly adding a random bond to the lattice,
identifying the clusters to which the sites at its ends belong by
traversing their respective trees until we find the root nodes, and then,
if necessary, amalgamating the two trees.  Generically, this kind of
algorithm is known as a ``union--find'' algorithm in the computer science
literature.  Our implementation for the percolation problem uses the
``weighted union--find with path compression''~\cite{Sedgewick88}, in which
(a)~two trees are always amalgamated by making the smaller a sub-tree of
the larger (``weighting'') and (b)~the pointers of all nodes along the path
traversed to reach the root node are changed to point directly to the root
(``path compression'').  Tarjan~\cite{Tarjan75} has shown for this
algorithm that the average number of steps taken to traverse the tree is
proportional to $\alpha(n)$, where $\alpha$ is the functional inverse of
Ackermann's function and $n$ is the number of nodes in the tree.  This in
turn implies that the number of steps is effectively constant as the tree
becomes large.  Our simulations confirm this result, the constant taking a
value of about $3.6$ on the square lattice, for example.  The computation
time taken in all other parts of the algorithm is also constant, and thus
the time taken for each bond to be added is $\O(1)$ and the time taken to
add all $N$ bonds is $\O(N)$.  Hence we can construct clusters for all
values of $n$ in one run of length~$\O(N)$.

The weighting in the tree union algorithm requires that we know the number
of nodes in each cluster.  This is easy to arrange however: we store the
cluster sizes at the root nodes of the clusters and when two clusters are
amalgamated we simply add their cluster sizes together.

In order to actually measure some quantity of interest we usually have to
do some additional work.  For example, if we wish to measure largest
cluster size we need to keep a running score of the largest cluster seen so
far, as the algorithm progresses.  If we wish to measure the position of
the percolation point, we can do so by adding variables to each site in a
cluster which store the displacement to the ``parent'' node in the tree.
Then when we traverse the tree, we add these displacements together to find
the total displacement to the root node.  When we add a bond to the lattice
which connects together two sites which belong to the same cluster, we
calculate the two such displacements and take their difference.  On a
lattice with periodic (toroidal) boundary conditions percolation has not
occurred if this difference is equal to a single lattice unit, otherwise it
has---see Fig.~\ref{tree}b.  (The same technique has been used to find the
percolation point of Fortuin--Kasteleyn clusters in the Potts
model~\cite{MCLSC96}.)


For site percolation, the algorithm is very similar to the one for the bond
case just described.  Sites are added in random order, and each one added
either forms a new detached cluster in its own right, joins onto a single
neighboring cluster, or joins together two or more extant clusters.  The
clusters are stored in a tree structure as before, and overall operation
takes time $\O(N)$.

We have tested our algorithm for both site and bond percolation on square
lattices of $L\times L$ sites with $L$ up to $10\,000$.  The time taken for
one run of our algorithm is found to scale as $N^\alpha$ with
$\alpha=1.02\pm0.04$, in agreement with the expected value of $\alpha=1$.
Even for the largest systems, a single run takes only about 100 seconds on
current computers.  Larger systems still would be easily within reach but
we are limited by the amount of memory available.  This is not an important
issue, however, since statistical error, rather than finite-size scaling,
is the principal factor limiting the accuracy of our numerical results,
making it more sensible to spend resources on reducing these errors than on
simulating especially large systems.

In this paper we apply our algorithm to the calculation of the probability
$R_L(p)$ for a cluster to wrap around the periodic boundary conditions on a
square lattice of $L\times L$ sites with site percolation.  For large $L$
this probability is equal to the probability that the system percolates.
Since cluster wrapping can be defined in a number of different ways there
are a corresponding number of different probabilities~$R_L$.  Here we
consider the following: $R_L^{(h)}$ and $R_L^{(v)}$ are the probabilities
of wrapping horizontally or vertically around the system respectively;
$R_L^{(b)}$ is the probability of wrapping around both directions
simultaneously; $R_L^{(e)}$ is the probability of wrapping around either
direction; and $R_L^{(1)}$ is the probability of wrapping around one
direction but not the other.  (Note that configurations which wrap around
both directions are taken to include both those which wrap directly around
the boundary conditions and ``spiral'' configurations in which a cluster
wraps around both directions before joining up.)  For the square systems
considered here, these probabilities satisfy the relations:
\begin{eqnarray}
R_L^{(h)} &=& R_L^{(v)},\nonumber\\
R_L^{(e)} &=& R_L^{(h)} + R_L^{(v)} - R_L^{(b)}
           =  2 R_L^{(h)} - R_L^{(b)},\\
R_L^{(1)} &=& R_L^{(h)} - R_L^{(b)} = R_L^{(e)} - R_L^{(h)}
           =  \mbox{$\frac12$} \bigl( R_L^{(e)} - R_L^{(b)} \bigr),\nonumber
\end{eqnarray}
as well as the inequalities $R_L^{(b)} \le R_L^{(h)} \le R_L^{(e)}$ and
$R_L^{(1)} \le R_L^{(h)}$.

Most previous studies of $R_L(p)$ have examined the probability of a
cluster connecting the boundaries of open systems.  The work presented
here differs from these studies by focusing on wrapping probabilities
for periodic systems, and constitutes the first precise such study.
As we will show, the use of wrapping probabilities yields estimates of
the position of the percolation threshold with much smaller
finite-size corrections than open-boundary methods.

\begin{figure}
\begin{center}
\psfig{figure=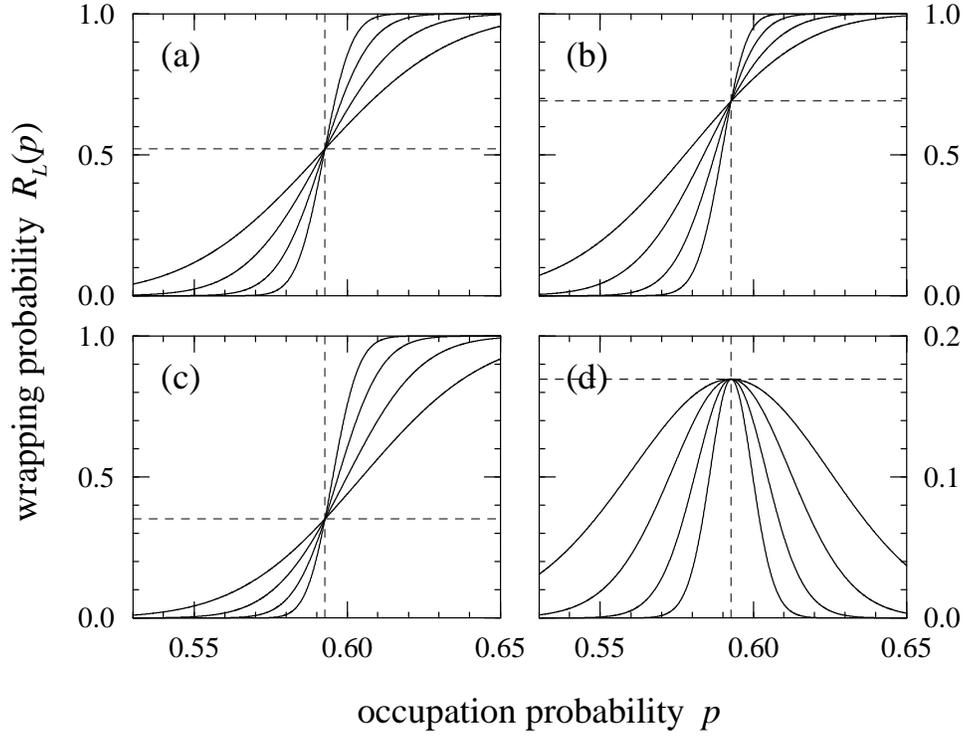,width=5in}
\end{center}
\caption{Plots of the cluster wrapping probability functions $R_L(p)$ for
  $L=32$, $64$, $128$ and $256$ in the region of the percolation transition
  for percolation (a)~along a specified axis, (b)~along either axis,
  (c)~along both axes, and (d)~along one axis but not the other.  Note that
  (d) has a vertical scale different from the other frames.  The dotted
  lines denote the expected values of $p_c$ and $R_\infty(p_c)$.}
\label{rl}
\end{figure}

To measure the wrapping probability we perform a number of runs of the
algorithm to find the number of occupied sites $n$ for which cluster
wrapping first occurs in the appropriate direction.  Then the corresponding
$R_L$ within the microcanonical ensemble is simply the fraction of runs for
which that point falls below~$n$.  Convolving the resulting curve with a
binomial according to Eq.~\eref{convolve} then gives us $R_L$ within the
canonical distribution.  Figure~\ref{rl} shows $R_L$ for each of the four
definitions above for a variety of system sizes.  Note that $R_L^{(1)}$ in
frame~(d) is non-monotonic, since the probability of wrapping around one
direction but not the other tends to zero as $p\to1$.

The exact values of $R_L$ at percolation for each of the definitions above
have been derived by Pinson~\cite{Pinson94,ZLK99}, and are
$R_\infty^{(h)}(p_c) = 0.521058290$, $R_\infty^{(e)}(p_c) = 0.690473725$,
$R_\infty^{(b)}(p_c) = 0.351642855$, and $R_\infty^{(1)}(p_c) =
0.169415435$.  We can use these figures to measure the value of $p_c$,
which is not known exactly for site percolation on the square lattice, by
finding the value of $p$ for which $R_L(p)=R_\infty(p_c)$.  These estimates
turn out to scale particularly well with system size.  For each of the
definitions of $R_L$ we find numerically that the difference
$R_L(p_c)-R_\infty(p_c)$ scales approximately as $L^{-2}$.  Since the width
of the critical region scales as $L^{-1/\nu}$, this implies that our
estimates of $p_c$ in finite systems should have a leading order
finite-size correction which goes as $L^{-2-1/\nu} = L^{-11/4}$.  This
represents a very rapid convergence, in contrast to the $L^{-1/\nu}$
behavior of typical percolation estimates (such as RG estimates) and the
$L^{-1-1/\nu}$ of certain open-system estimates~\cite{Ziff92}, which is the
best previously known convergence.  Note that if the microcanonical values
of $R_L$ are used instead of the canonical ones, the difference
$R_L(p_c)-R_\infty(p_c)$ scales as $L^{-1/2}$, making this method
significantly inferior to the one described above.

The non-monotonic probability function $R_L^{(1)}(p)$ is never equal to
$R_\infty^{(1)}(p_c)$ because the value of $R_L^{(1)}$ on systems of finite
size is less than the value at $L=\infty$.  However, in this case we can
estimate $p_c$ from the position of the maximum of the function, and this
estimate is also expected to scale as $L^{-2-1/\nu}$.

\begin{figure}
\begin{center}
\psfig{figure=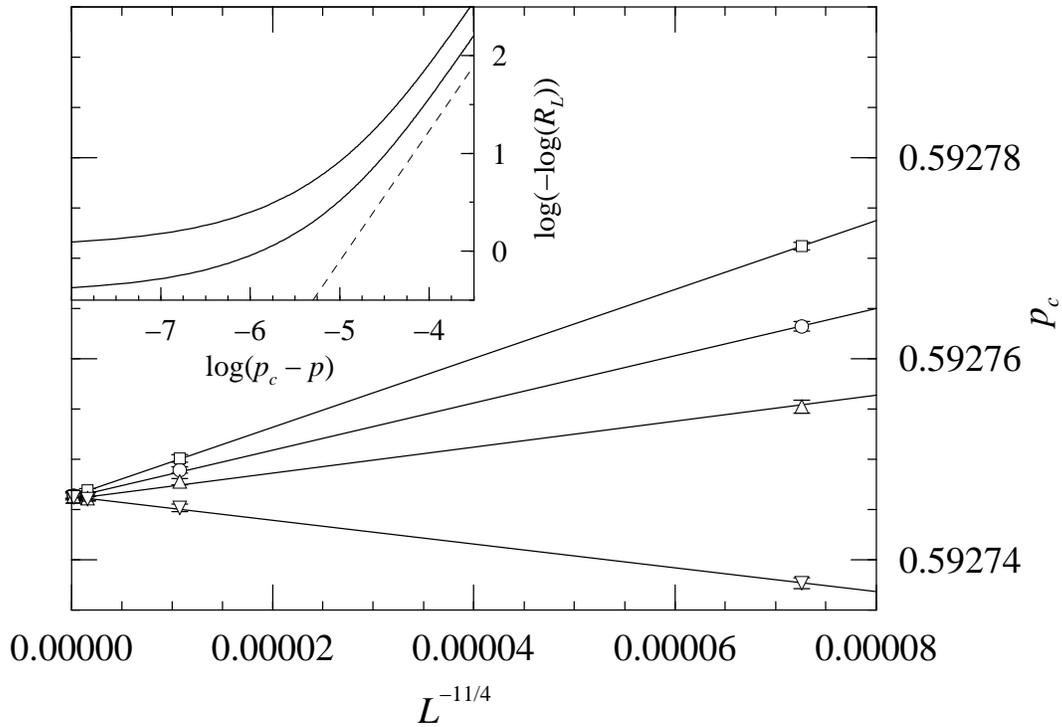,width=5in}
\end{center}
\caption{Finite-size scaling of the estimated value of $p_c$ for site
  percolation for $R_L^{(h)}$ (circles), $R_L^{(e)}$ (squares), $R_L^{(b)}$
  (upward-pointing triangles), and $R_L^{(1)}$ (downward-pointing
  triangles).  Inset: scaling plot of the wrapping probabilities
  $R_L^{(h)}$ (lower curve) and $R_L^{(b)}$ (upper curve).  The dotted line
  indicates the slope of the expected $\frac43$-power tail.}
\label{scaling}
\end{figure}

In Fig.~\ref{scaling} we show the values of $p_c$ estimated from our Monte
Carlo results as a function of $L^{-11/4}$ for $L=32$, $64$, $128$, and
$256$ for each of the four definitions of $R_L$.  At least $3\times10^8$
runs were carried out for each system size to achieve high statistical
accuracy.  Two different random number generators were used: a two-tap
32-bit additive lagged Fibonacci generator with taps at 418 and 1279, and a
four-tap 32-bit XOR generator with taps at 471, 1586, 6988 and 9689.
Allowing for statistical fluctuation, results were consistent between
generators.

The best fits to $L^{-11/4}$ give estimates for the position of the
percolation threshold for site percolation on the square lattice of
$0.59274621(13)$ for $R_L^{(h)}$, $0.59274636(14)$ for $R_L^{(e)}$,
$0.59274606(15)$ for $R_L^{(b)}$, and $0.59274629(20)$ for $R_L^{(1)}$.
Our best estimate of $p_c$ is therefore
\begin{equation}
p_c = 0.59274621 \pm 0.00000013,
\end{equation}
which is more accurate by a factor of four than the best previously
published estimate of this quantity~\cite{Ziff92,note2} and should prove
useful for high-precision studies of percolation in the future.

While it is encouraging to be able to estimate $p_c$ so accurately, the
real power of our algorithm lies in its ability to efficiently estimate a
function such as $R_L(p)$ over the entire range of $p$.  To demonstrate the
application of this idea, we have used our simulations to extract explicit
evidence of the expected $\frac43$-power tail in the logarithm of the
cluster wrapping probability function.

The probability $R_L(p)$ that a given realization of a percolation model
will wrap around a finite lattice at a particular value of $p$ is expected
to go as $\exp(-L/\xi)$ when $\xi\ll L$~\cite{HA96,BW95}.  Putting
$\xi\sim(p_c-p)^{-\nu}$ we thus get
\begin{equation}
R_L \sim \exp(-L (p_c-p)^\nu).
\label{tails}
\end{equation}
This variation with $p$ is difficult to detect using standard algorithms
for measuring $R_L$ (see, for example, Ref.~\onlinecite{HA96}), since one
needs to generate large numbers of samples at many different values of $p$,
and almost all of that work will be wasted, since most of the systems
simulated do not percolate.  Our algorithm however shows the behavior of
Eq.~\eref{tails} clearly without further work.  Equation~\eref{tails}
implies that a plot of $\log(-\log R_L)$ against $\log(p_c-p)$ should have
an asymptotic slope of $\nu=\frac43$.  In the inset of Fig.~\ref{scaling}
we show such a plot for the functions $R_L^{(h)}$ and $R_L^{(b)}$ for
systems with $L=256$.  The $\frac43$-power tail is clearly visible.

To conclude, we have presented a new Monte Carlo algorithm for studying
site or bond percolation on any lattice.  The algorithm is capable of
measuring the entire curve of an observable quantity as a function of the
occupation probability $p$ in a single run taking time of order the volume
of the system.  We have also proposed a new and highly accurate method for
measuring the position of the percolation threshold by calculating the
probabilities for clusters to wrap around the boundary conditions on a
toroidal system.  We have used this method in combination with our Monte
Carlo algorithm to find the value of $p_c$ for site percolation on the
square lattice to greater accuracy than any previously published
calculation.  In addition we have used our algorithm to demonstrate clearly
the presence of the expected $\frac43$-power tails in the logarithm of the
cluster wrapping probability.

The authors would like to thank Harvey Gould, Cris Moore, and Barak
Pearlmutter for helpful comments.

\newpage

\newpage

\end{document}